\documentclass{article}
\usepackage{color}
\usepackage{booktabs}
\usepackage{amsfonts}
\usepackage{spconf,amsmath,graphicx}
\usepackage{siunitx}
\usepackage{tikz}
\usepackage{pgfplotstable}
\usepackage{subfig}
\usepackage{hyperref}
\usepackage{float}
\let\OLDthebibliography\thebibliography
\renewcommand\thebibliography[1]{
  \OLDthebibliography{#1}
  \setlength{\parskip}{0pt}
  \setlength{\itemsep}{0pt plus 0.5ex}
}
\pgfplotsset{compat=1.14}

\title{Meta Learning for End-to-End Low-Resource Speech Recognition}
\name{Jui-Yang Hsu\qquad Yuan-Jui Chen\qquad  Hung-yi Lee}
\address{National Taiwan University \\College of Electrical Engineering and Computer Science \\ 
\small{\texttt{\{r07921053, r07922070, hungyilee\}@ntu.edu.tw}}}
\begin{document}
%\ninept
%
\maketitle
\begin{abstract}
  In this paper, we proposed to apply meta learning approach for low-resource automatic speech recognition (ASR). We formulated ASR for different languages as different tasks, and meta-learned the initialization parameters from many pretraining languages to achieve fast adaptation on unseen target language, via recently proposed model-agnostic meta learning algorithm (MAML). We evaluated the proposed approach using six languages as pretraining tasks and four languages as target tasks. Preliminary results showed that the proposed method, MetaASR, significantly outperforms the state-of-the-art multitask pretraining approach on all target languages with different combinations of pretraining languages. In addition, since MAML's model-agnostic property, this paper also opens new research direction of applying meta learning to more speech-related applications.
\end{abstract}

\section{Introduction}
\label{sec:intro}

With the recent advances of deep learning, integrating the main modules of automatic speech recognition (ASR) such as acoustic model, pronunciation lexicon, and language model into a single end-to-end model is highly attractive. Connectionist Temporal Classification (CTC) \cite{graves2006connectionist} lends itself on such end-to-end approaches by introducing an additional blank symbol and specifically-designed loss function optimizing to generate the correct character sequences from the speech signal directly, without framewise phoneme alignment in advance. With many recent results \cite{hannun2014deep, amodei2016deep, collobert2016wav2letter}, end-to-end deep learning has created a larger interest in the speech community.

However, to build such an end-to-end ASR system requires a huge amount of paired speech-transcription data, which is costly. 
For most languages in the world, they lack sufficient paired data for training. 
Pretraining on other language sources as the initialization, then fine-tuning on target language is the dominant approach under such low-resource setting, also known as multilingual transfer learning / pretraining (MultiASR) \cite{vu2014multilingual, tong2017investigation}. 
The backbone of MultiASR is a multitask model with shared hidden layers (encoder), and many language-specific heads. 
The model structure is designed to learn an encoder to extract language-independent representations to build a better acoustic model from many source languages. 
The success of ``language-independent'' features to improve ASR performance compared to monolingual training has been shown in many recent works \cite{cho2018multilingual, dalmia2018sequence, tong2017multilingual}.

\begin{figure}[htb]

\begin{minipage}[b]{0.48\linewidth}
  \centering
  \centerline{\includegraphics[width=4.0cm]{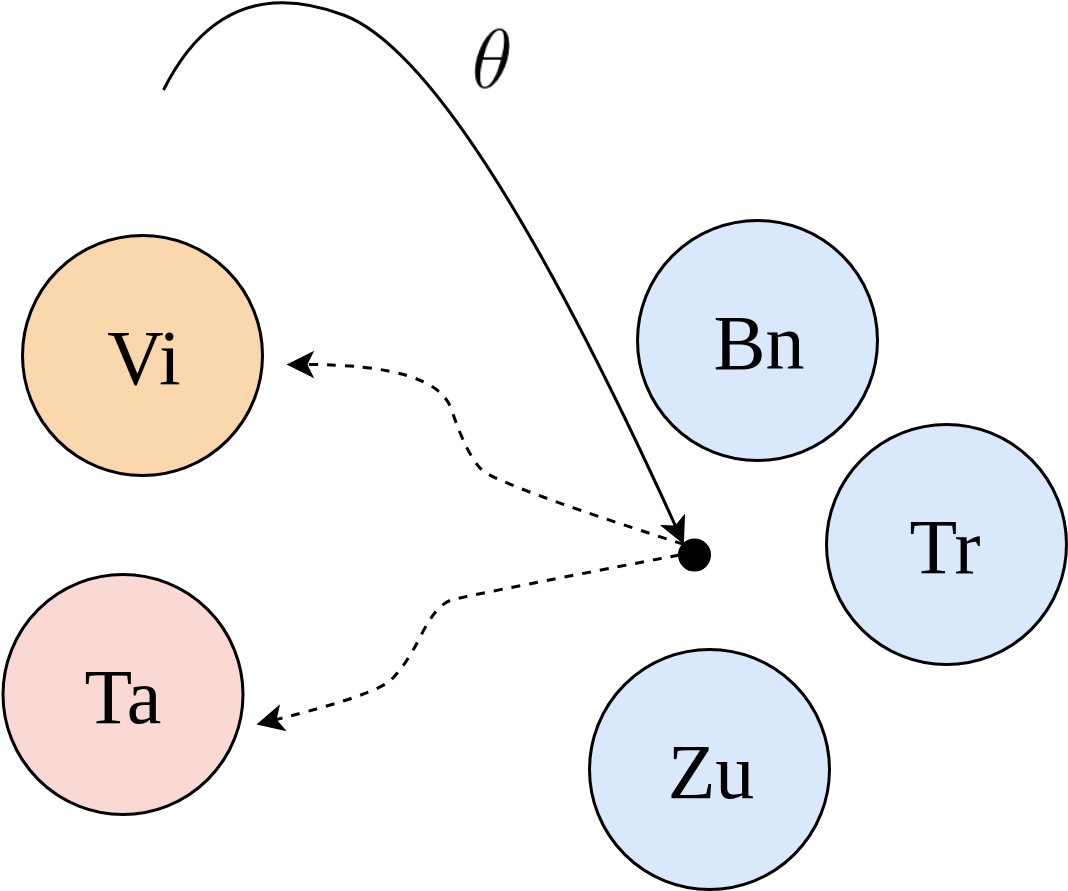}}
%  \vspace{1.5cm}
  \centerline{(a) MultiASR}\medskip
\end{minipage}
\hfill
\begin{minipage}[b]{0.48\linewidth}
  \centering
  \centerline{\includegraphics[width=4.0cm]{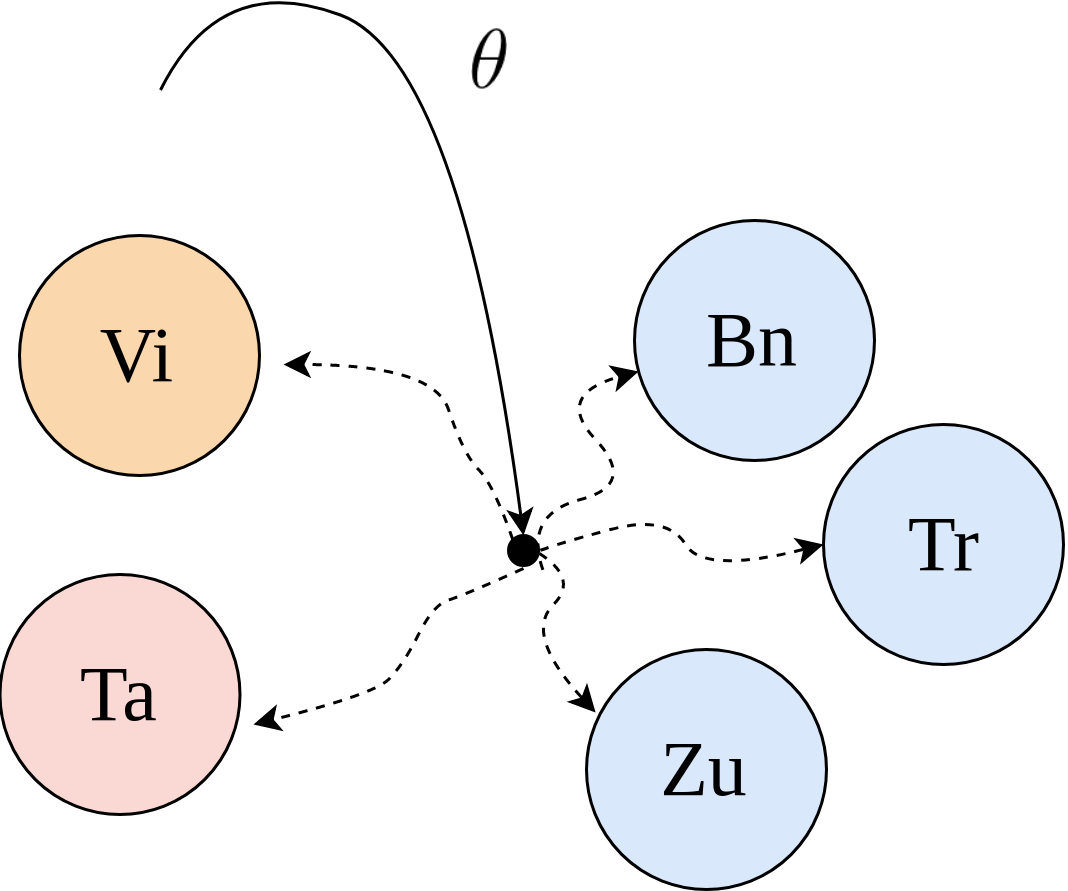}}
%  \vspace{1.5cm}
  \centerline{(b) MetaASR}\medskip
\end{minipage}
\caption{Illustration: Difference of the learned parameters from MultiASR \& MetaASR. The solid lines represent the learning process of pretraining, either multitask or meta learning. The dashed lines represent the language-specific adaptation.\\ (The figure is modified from \cite{gu2018meta})}
\label{fig:meta-idea}
\end{figure}

Besides directly training the model with all the source languages, there are various variants of MultiASR approaches. 
Language-adversarial training approaches \cite{Yi2018AdversarialMT, adams2019massively} introduced language-adversarial classification objective to the shared encoder, negating the gradients backpropagated from the language classifier to encourage the encoder to extract more language-independent representations. 
Hierarchical approaches \cite{Sanabria2018HierarchicalMT} introduced different granularity objectives by combining both character and phoneme prediction at different levels of the model.

%Lee: 我暫時把兩段合併，這邊應該要提到說，我們用的方法是 MAML ，可以跟任何 network architecture 結合
In this paper, we provide a novel research direction following up on the idea of multilingual pretraining -- \textbf{Meta learning}. 
Meta learning, or learning-to-learn, has recently received considerable interest in the machine learning community. The goal of meta learning is to solve the problem of ``fast adaptation on unseen data'', which is aligned with our low-resource setting.
 With its success in computer vision under the few-shot learning setting~\cite{rusu2018meta, snell2017prototypical, vinyals2016matching}, there have been some works in language and speech processing, for instance, language transfer in neural machine translation \cite{gu2018meta}, dialogue generation \cite{mi2019meta}, and speaker adaptive training \cite{klejch2018learning}, but not multilingual pretraining for speech recognition.

We use model-agnostic meta-learning algorithm (MAML) \cite{finn2017model} in this work. 
As its name suggestes, MAML can be applied to any network architecture. 
MAML only modifies the optimization process following meta learning training scheme.
It does not introduce additional modules like adversarial training or requires phoneme level annotation (usually through lexicon) such as hierarchical approaches. 
We evaluated the effectiveness of the proposed meta learning algorithm, MetaASR, on the IARPA BABEL dataset \cite{gales2014speech}. Our experiments reveal that MetaASR outperforms MultiASR significantly across all target languages.
\section{Proposed Approach}
\label{sec:approach}

\subsection{Multilingual CTC Model}
\begin{figure}[t]
    \centering
    \includegraphics[width=0.8\linewidth]{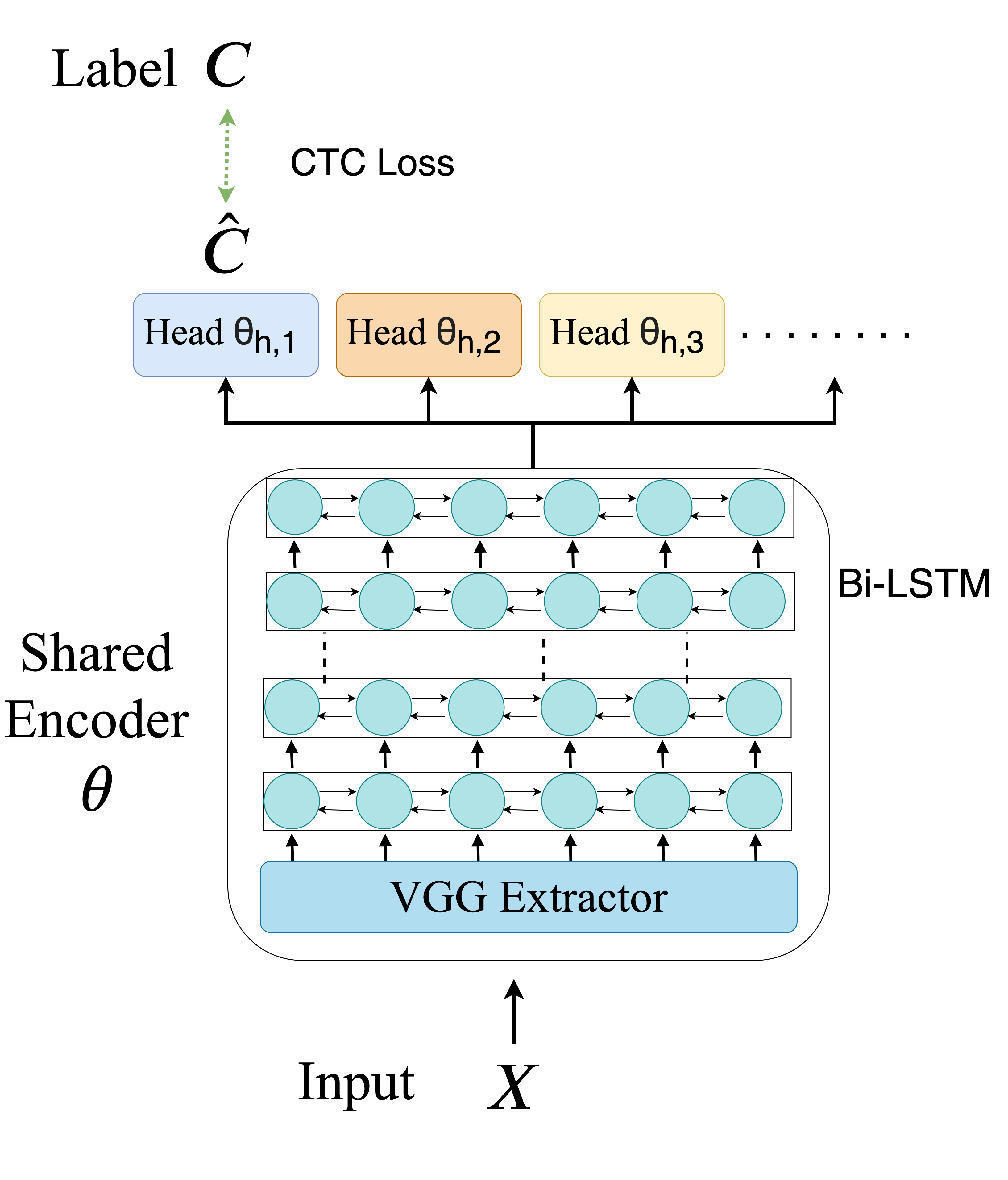}
    \caption{Multilingual CTC model architecture}
    \label{fig:model-arch}
\end{figure}

%We used the model architecture as illustrated in Fig.~\ref{fig:model-arch}, the shared encoder is parameterized by $\theta$, and the set of language-specific heads are parameterized by $\theta_h$ ($\theta_{h,l}$ means the head for $l$-th language). 
We used the model architecture as illustrated in Fig.~\ref{fig:model-arch}, the shared encoder is parameterized by $\theta$, and the set of language-specific heads are parameterized by $\theta_{h,l}$ (the head for $l$-th language). %LeeNew:這樣會不會比較好 ?
Let the dataset be $D$, composed of paired data $(X,C)$. Let $X = x_1, x_2, \cdots, x_T$ with length $T$ as input feature, $C = c_1, c_2, \cdots, c_L$ with length $L$ as target label. $X$ is encoded into sequence of hidden states through the shared encoder, then fed into the language-specific head of the corresponding language with softmax activation to output the prediction sequence $\hat{C} = \hat{c}_1, \hat{c}_2, \cdots, \hat{c}_{L^\prime}$ with length $L^\prime$.

%\vspace{20pt}

\textbf{CTC Loss}. CTC computes the posterior probability as below,

\begin{equation}
  P(C|X) = \sum_{\pi \in \mathcal{Z}(C)} P(\pi|X)
\end{equation}
where $\pi$ is the repeated character sequence  of $C$ with additional blank label, and $\mathcal{Z}(C)$ is the set of all possible sequences given $C$. For each $\pi$, we can approximate the posterior probability as below,

\begin{equation}
  P(\pi|X) \approx \prod_{i=1}^{L^\prime} P(\hat{c_i}|X)
\end{equation}

Take X belonging to the $l$-th language for instance, the loss function of the model on $D$ is then defined as:

\begin{equation}
  \label{eq:ctc-loss}
  \mathcal{L}_D(\theta, \theta_{h,l}) = - \log P(C|X)
\end{equation}

\subsection{Meta Learning for Low-Resource ASR}

The idea of MAML is to learn initialization parameters from a set of tasks. 
In the context of ASR, we can view different languages as different tasks.
Given a set of source tasks $\mathcal{D}=\lbrace D_1, D_2, \cdots, D_K \rbrace$, MAML learns from $\mathcal{D}$ to obtain good initialization parameters $\theta^{\star}$ for the shared encoder.
$\theta^{\star}$ yields fast task-specific learning (fine-tuning) on target task $D_t$ and obtains $\theta^{\star}_t$ and $\theta^{\star}_{h,t}$ (the parameters obtained after fine-tuning on $D_t$). 
MAML can be formulated as below, 
\begin{equation*}
  \theta^{\star}_t, \theta^{\star}_{h,t} = \texttt{Learn}(D_t;\theta^{\star}) = \texttt{Learn}(D_t;\texttt{MetaLearn}(\mathcal{D})).
\end{equation*}
The two functions,  \texttt{Learn} and \texttt{MetaLearn}, will be described in the following two subsections.
%where $\theta^{\star}_t$ is the parameter obtained after fine-tuning on $T_t$.

\subsubsection{Learn: Language-specific learning}
Given any initial parameters $\theta^0$ of the shared encoder (either random initialized or obtained from pretrained model) and the dataset $D_t$. The language-specific learning process is to minimize the CTC loss function defined in Eq.~\ref{eq:ctc-loss}.

\begin{equation}
  \label{eq:fine-tune}
  \resizebox{0.91\hsize}{!}{$
\begin{aligned}
  \theta^\prime, \theta^\prime_{h,t} = \texttt{Learn}(D_t;\theta^0) & = \arg \, \min_{\theta, \theta_{h,t}} \mathcal{L}_{D_t}(\theta, \theta_{h,t}) \\
                                                                    & = \arg \, \min_{\theta, \theta_{h,t}} \sum_{(X,C) \in D_t} -\log P(C|X)
\end{aligned}
$}
\end{equation}
\begin{table*}[thp!]
\centering
\caption{Character error rate (\si{\percent} CER) w.r.t the pretraining languages set for all 4 target languages' FLP}
\label{tab:flp-table}
\begin{tabular}{@{}ccccccccc@{}}
%\begin{tabular}{l|cc|cc|cc|cc}
\toprule
Model                                    & \multicolumn{2}{c}{Vietnamese}                         & \multicolumn{2}{c}{Swahili}                        & \multicolumn{2}{c}{Tamil}                        & \multicolumn{2}{c}{Kurmanji} \\

                                         & multi           & meta                                & multi           & meta                                & multi           & meta                                & multi           & meta           \\ \midrule
\multicolumn{1}{c|}{(no-pretrain)}                   & \multicolumn{2}{c|}{71.8}                    & \multicolumn{2}{c|}{47.5}                    & \multicolumn{2}{c|}{69.9}          & \multicolumn{2}{c}{64.3}                    \\

\multicolumn{1}{l|}{Bn Tl Zu}   &    57.4      & \multicolumn{1}{c|}{49.9}          & 48.1          & \multicolumn{1}{c|}{41.4}          & 65.6          & \multicolumn{1}{c|}{57.5}          & 61.1          & 57.0          \\
\multicolumn{1}{l|}{ \qquad \qquad Tr Lt Gn} & 63.7          & \multicolumn{1}{c|}{49.5}          & 57.2          & \multicolumn{1}{c|}{41.8}          & 68.2          & \multicolumn{1}{c|}{57.7}          & 65.6          & 57.0          \\
\multicolumn{1}{l|}{Bn Tl Zu Tr Lt Gn}           & 59.7          & \multicolumn{1}{c|}{50.1}          & 48.8          & \multicolumn{1}{c|}{42.9}          & 65.6          & \multicolumn{1}{c|}{58.9}          & 62.6          & 57.6          \\ \bottomrule
\end{tabular}
\end{table*}

The learning process is optimized through gradient descent, the same as MultiASR.

\subsubsection{MetaLearn}
The initialization parameters found by MAML should not only adapt to one language well, but for as many languages as possible. To achieve this goal, we define the meta learning process and the corresponding meta-objective as follows.

In each meta learning episode, we sample batch of tasks from $\mathcal{D}$, then sample two subsets from each task $k$ as training and testing set, denoted as $D_{k}^{tr}$, $D_{k}^{te}$, respectively. First, we use $D_k^{tr}$ to simulate the language-specific learning process to obtain $\theta^\prime_k$ and $\theta^\prime_{h,k}$.

\begin{equation}
    \theta_k^\prime, \theta_{h,k}^\prime = \texttt{Learn}(D^{tr}_k; \theta)
                                         %&= \arg \, \min_{\theta, \theta_{h,k}} \mathcal{L}_{D^{te}_k}(\theta, \theta_{h,k})
\end{equation}

Then evaluate the effectiveness of the obtained parameters on $D_k^{te}$. The goal of MAML is to find $\theta$, the initialization weights of the shared encoder for fast adaptation, so the meta-objective is defined as

\begin{equation}
  \label{eq:meta-obj}
  \mathcal{L}^{\text{meta}}_{\mathcal{D}}(\theta) =  \mathbb{E}_{k \sim \mathcal{D}} \; \mathbb{E}_{D_k^{tr}, D_k^{te}} \Big [ \mathcal{L}_{D^{te}_k}(\theta^\prime_k, \theta^\prime_{h,k}) \Big ]
\end{equation}

Therefore, the meta learning process is to minimize the loss function defined in Eq.~\ref{eq:meta-obj}.

\begin{equation}
  \theta^\star = \texttt{MetaLearn}(\mathcal{D}) = \arg \min_{\theta} \mathcal{L}^{\text{meta}}_{\mathcal{D}} (\theta)
\end{equation}

We use \textit{meta gradient} obtained from Eq.~\ref{eq:meta-obj} to update the model through gradient descent.

\begin{equation}
  \label{eq:meta-grad}
  \theta \leftarrow \theta - \eta^\prime \sum_k \nabla_\theta \mathcal{L}_{D^{te}_k}(\theta^\prime_k, \theta^\prime_{h,k})
\end{equation}
$\eta^\prime$ is the meta learning rate. And noted that only the shared encoder is updated via Eq.~\ref{eq:meta-grad}.

MultiASR optimizes the model according to Eq.~\ref{eq:fine-tune} on all source languages directly, without considering how learning happens on the unseen language. Although the parameters found by MultiASR is good for all source languages, it may not adapt well on the target language. Unlike MultiASR, MetaASR explicitly integrates the learning process into its framework via simulating language-specific learning first, then meta-updates the model. Therefore, the parameters obtained are more suitable to adapt on the unseen language. We illustrate the concept in Fig.~\ref{fig:meta-idea}, and show it in the experimental results in Section \ref{sec:results}.

\section{Experiment}
\label{sec:exp}

%Lee:我覺得這章太多節了

\label{ssec:exp-setup}
In this work, we used data from the IARPA BABEL project \cite{gales2014speech}. 
The corpus is mainly composed of conversational telephone speech (CTS). We selected 6 languages as non-target languages for multilingual pretraining: 
Bengali (Bn), Tagalog (Tl), Zulu (Zu), Turkish (Tr), Lithuanian (Lt), Guarani (Gn), and 4 target languages for adaptation: Vietnamese (Vi), Swahili (Sw), Tamil (Ta), Kurmanji (Ku), and experimented different combinations of non-target languages for pretraining.
Each language has Full Language Pack (FLP) and Limited Language Pack (LLP, which consists of 10\% of FLP).
%這裡解釋 FLP LLP

We followed the recipe provided by Espnet \cite{watanabe2018espnet} for data preprocessing and final score evaluation.  We used 80-dimensional Mel-filterbank and 3-dimensional pitch features as acoustic features. The size of the sliding window is 25ms, and the stride is 10ms. We used the shared encoder with a 6-layer VGG extractor with downsampling and a 6-layer bidirectional LSTM network with 360 cells in each direction as used in the previous work \cite{dalmia2018sequence}.

\textbf{Meta Learning}. For each episode, we used a single gradient step of language-specific learning with SGD when computing the meta gradient. Noted that in Eq.~\ref{eq:meta-grad}, if we expanded the loss term in the summation via Eq.~\ref{eq:fine-tune}, we would find the second-order derivative of $\theta$ appear. For computation efficiency, some previous works \cite{finn2017model, nichol2018reptile} showed that we could ignore the second-order term without affecting the performance too much.\\
Therefore, we approximated Eq.~\ref{eq:meta-grad} as follows.

\begin{equation}
  \theta \leftarrow \theta - \eta^\prime \sum_k \nabla_{\textcolor{red}{\theta_k^\prime}} \mathcal{L}_{D^{te}_k}(\theta^\prime_k, \theta^\prime_{h,k})
\end{equation}
Also known as First-order MAML (FOMAML).

We multi-lingually pretrained the model for 100K steps for both MultiASR and MetaASR. When adapting to one certain language, we used the LLP of the other three languages as validation sets to decide which pretraining step we should pick. Then we fine-tuned the model 18 epochs for the target language on its FLP, 20 epochs on its LLP, and evaluated the performance on the test set via beam search decoding with beam size $20$ and 5-gram language model re-scoring,  as Table~\ref{tab:flp-table} and \ref{tab:llp-table} displayed.

\section{Results}
\label{sec:results}
\begin{table*}[ht!]
\centering
\caption{Character error rate (\si{\percent} CER) w.r.t the pretraining languages set for all 4 target languages' LLP}
\label{tab:llp-table}
\begin{tabular}{@{}ccccccccc@{}}
%\begin{tabular}{l|cc|cc|cc|cc}
\toprule
Model                                    & \multicolumn{2}{c}{Vietnamese}                         & \multicolumn{2}{c}{Swahili}                        & \multicolumn{2}{c}{Tamil}                        & \multicolumn{2}{c}{Kurmanji} \\

                                         & multi           & meta                                & multi           & meta                                & multi           & meta                                & multi           & meta           \\ \midrule
\multicolumn{1}{c|}{(no-pretrain)}                   & \multicolumn{2}{c|}{74.7}                    & \multicolumn{2}{c|}{65.0}                    & \multicolumn{2}{c|}{72.4}          & \multicolumn{2}{c}{68.9}                    \\

\multicolumn{1}{l|}{Bn Tl Zu}   & 65.0          & \multicolumn{1}{c|}{58.1}          & 62.6          & \multicolumn{1}{c|}{57.5}          & 70.4          & \multicolumn{1}{c|}{73.7}          & 67.6         & 64.6          \\
\multicolumn{1}{l|}{ \qquad \qquad Tr Lt Gn} & 64.9          & \multicolumn{1}{c|}{58.0}          & 64.1          & \multicolumn{1}{c|}{59.6}          & 73.7          & \multicolumn{1}{c|}{74.7}          & 69.7          & 63.0          \\
\multicolumn{1}{l|}{Bn Tl Zu Tr Lt Gn}           & 64.1          & \multicolumn{1}{c|}{58.7}          & 61.9          & \multicolumn{1}{c|}{59.6}          & 70.0          & \multicolumn{1}{c|}{68.2}          & 66.7          & 64.1          \\ \bottomrule
\end{tabular}
\end{table*}

%\subsection{Performance Comparison: CER on FLP}
\label{ssec:multitask-baseline}

%\subsection{Performance Comparison of CER on FLP}
\textbf{Performance Comparison of CER on FLP}. As presented in Table~\ref{tab:flp-table}, compared to monolingual training (that is, without using pretrained parameters as initialization, denoted as no-pretrain), both MultiASR and MetaASR improved the ASR performance using different combinations of pretraining languages. Table~\ref{tab:flp-table} clearly shows that the proposed MetaASR significantly outperforms MultiASR across all target languages. We were also interested in the impact of the choices of pretraining languages and found that the performance variance of MetaASR is smaller than MultiASR. It might be due to the fact that MetaASR focuses more on the learning process rather than fitting on source languages. 

\begin{figure}[ht]
  \centering
  %\hspace{-2.2cm}
  %\begin{tikzpicture}[trim axis left, trim axis right]
  \begin{tikzpicture}

  \begin{axis}[
    width=\linewidth,
    height=6.0cm,
    legend entries={MultiASR, MetaASR, no-pretrain} ,
    xlabel = {Number of pretraining steps ($\times 1000$)},
        xmin=5,
        ymin=36,
        ymax=67,
        grid=both,
        legend style={at={(0.02,0.81)},anchor=north west},
        %legend pos=outer north east,
        ylabel={CER (\si{\percent}})]
  \addplot+[smooth]table{multi-stat/multi3-swahili};
  \addplot+[smooth]table{meta-stat/meta3-swahili};
   \addplot[style=ultra thick,dashed,] coordinates {(0,64.3) (100,64.3)};

  \end{axis}
  \end{tikzpicture}
  \caption{Learning curves on Swahili's LLP \\ pretrained on Bn, Tl, Zu}
  \label{fig:curve1}
\end{figure}
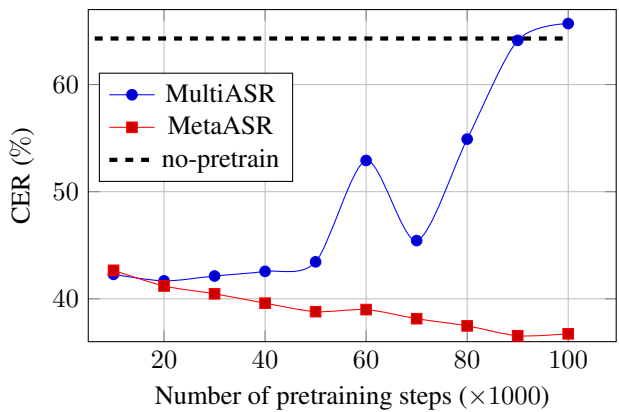
\begin{figure}[ht]
  \centering
  %\hspace{-2.2cm}
  \begin{tikzpicture}
  %\begin{tikzpicture}[trim axis left, trim axis right]

  \begin{axis}[
    width=\linewidth,
    height=6.0cm,
    legend entries={MultiASR, MetaASR, no-pretrain} ,
    xlabel = {Number of pretraining steps ($\times 1000$)},
        xmin=5,
        grid=both,
        legend style={at={(0.02,0.48)},anchor=south west},
        %legend pos=inner north west,
        ylabel={CER (\si{\percent}})]
  \addplot+[smooth]table{multi-stat/multi6-swahili};
  \addplot+[smooth]table{meta-stat/meta6-swahili};
   \addplot[style=ultra thick,dashed,] coordinates {(0,64.3) (100,64.3)};
  \end{axis}
  \end{tikzpicture}
  \center \caption{Learning curves on Swahili's LLP \\  pretrained on Bn, Tl, Zu, Tr, Lt, Gn}
  \label{fig:curve2}
\end{figure}
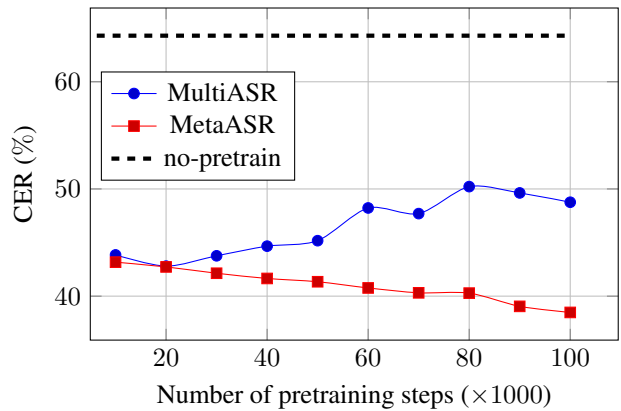

\noindent
%\subsection{Training Curves}
\textbf{Learning Curves}. The advantage of MetaASR over MultiASR is clearly shown in Fig.~\ref{fig:curve1} and \ref{fig:curve2}. Given the pretrained parameters of the specific pretraining step, we fine-tuned the model for 20 epochs and reported the lowest CER on its validation set. The above process represented one point of the curve. For MultiASR, the performance of adaptation saturated in the early stage and finally degraded.  As Fig.~\ref{fig:meta-idea} illustrates, the training scheme of MultiASR tended to overfit on pretraining languages, and the learned parameters might not be suitable for adaptation. From Fig.~\ref{fig:curve1}, we can see that in the later stage of pretraining, using such pretrained weights even yields worse performance than random initialization. In contrast, for MetaASR, not only the performance is better than MultiASR during the whole pretraining process, but it also gradually improves as pretraining continues without degrading.  The adaptation of all languages using different pretraining languages show similar trends.  We only showed the results of Swahili here due to space limitations.
%LeeNew: 我改了一下這句

%\subsection{Impact on Training Set Size}
~\\
\noindent
\textbf{Impact on Training Set Size}. In addition to adapting on FLP of the target languages, we have also fine-tuned on LLP of them, and the result is shown in Table~\ref{tab:llp-table}. On Vietnamese, Swahili, and Kurmanji, MetaASR also outperforms MultiASR. Both of MultiASR and MetaASR improve the performance, but the gap compared to the no-pretrain model is smaller than fine-tuning on FLP. 
On Tamil, weights from pretrained model was even worse than random initialization. We will evaluate more combinations of target languages and pretraining languages to investigate the potential of our proposed method in such ultra low-resource scenario.

\section{Conclusion}
\label{sec:conclusion}
In this paper, we proposed a meta learning approach to multilingual pretraining for speech recognition. The initial experimental results showed its potential in multilingual pretraining. In future work, we plan to use more combinations of languages and corpora to evaluate the effectiveness of MetaASR extensively. Besides, based on MAML's model-agnostic property, this approach can be applied to a wide range of network architectures such as sequence-to-sequence model, and even different applications beyond speech recognition.

% Below is an example of how to insert images. Delete the ``\vspace'' line,
% uncomment the preceding line ``\centerline...'' and replace ``imageX.ps''
% with a suitable PostScript file name.
% -------------------------------------------------------------------------

%\vfill
%\pagebreak

%\section{REFERENCES}
\bibliographystyle{IEEEbib}
\bibliography{strings,refs}

% 以下不會放到 paper ，只是畫一些圖出來而已
\end{document}